# A New Switched Reluctance Motor with Embedded Permanent Magnets for Transportation Electrification

Gholamreza Davarpanah
*Department of Electrical Engineering*
Amirkabir University of Technology
Tehran, Iran
ghr.davarpanah@aut.ac.ir

Sajjad Mohammadi
*Department of Electrical Engineering and Computer Science*
*MIT*
Cambridge, USA
sajjadm@mit.edu

James L. Kirtley
*Department of Electrical Engineering and Computer Science*
*MIT*
Cambridge, USA
kirtley@mit.edu

***Abstract*—A new three-phase hybrid-excited multi-tooth switched reluctance motor with embedded permanent magnets is proposed, capable of achieving higher torque density for transportation electrification applications. Operating principles and design considerations of the proposed motor are discussed. The finite element method is employed in the field analysis and the design process. The advantages of the proposed topology over existing designs for switched reluctance motors and flux switching motors are presented. Finally, the optimized design is prototyped to experimentally confirm the design and simulation results.**

## I. Introduction

Recently, there has been increased interest in electric and hybrid electric vehicles, with a focus on improving the efficiency, size, weight, reliability, speed range, and cost of the drivetrain, including electric motors, power electronics drives, and battery systems. Switched reluctance motors (SRMs) have the potential to address these issues with robust and fault-tolerant drives for electric propulsion [1]-[8]. New SRM designs have been studied in [9]-[19]. Achieving higher torque is critical to improving the overall performance of SRMs [20]-[27]. Embedding permanent magnets (PMs) into SRMs has improved torque density [28]-[30], yet cogging torque remains inevitable [31]. A multi-tooth hybrid excited SRM (MT-HESRM) [32] and an SRM with two sets of PMs (PM-SRM) [33] have been introduced, but the unwanted radial forces are significant which can generate acoustic noise and negatively affect the long-term operation of the motor. A larger number of stator teeth has been adopted as mitigation for radial forces [34]-[35], but it results in narrower rotor poles, causing saturation, higher core losses, and a shorter commutation angle requiring a faster switching frequency. In [36], a modular SRM with an A-type stator is introduced, incorporating PMs within the stator back iron.

This paper proposes a three-phase hybrid-excited multi-tooth SRM (HEMTSRM) with embedded PMs to increase torque density, making it a potential choice for transportation electrification applications. At higher stator currents, the PMs contribute to energy conversion and torque production. The finite element method (FEM) is employed in the design process. The operating principles and magnetic fields of the proposed SRM are analyzed. A comparison is made with conventional SRMs as well as a flux switching permanent magnet motor. Finally, the optimized design is prototyped and experimentally tested to confirm the new design and the FEM results.

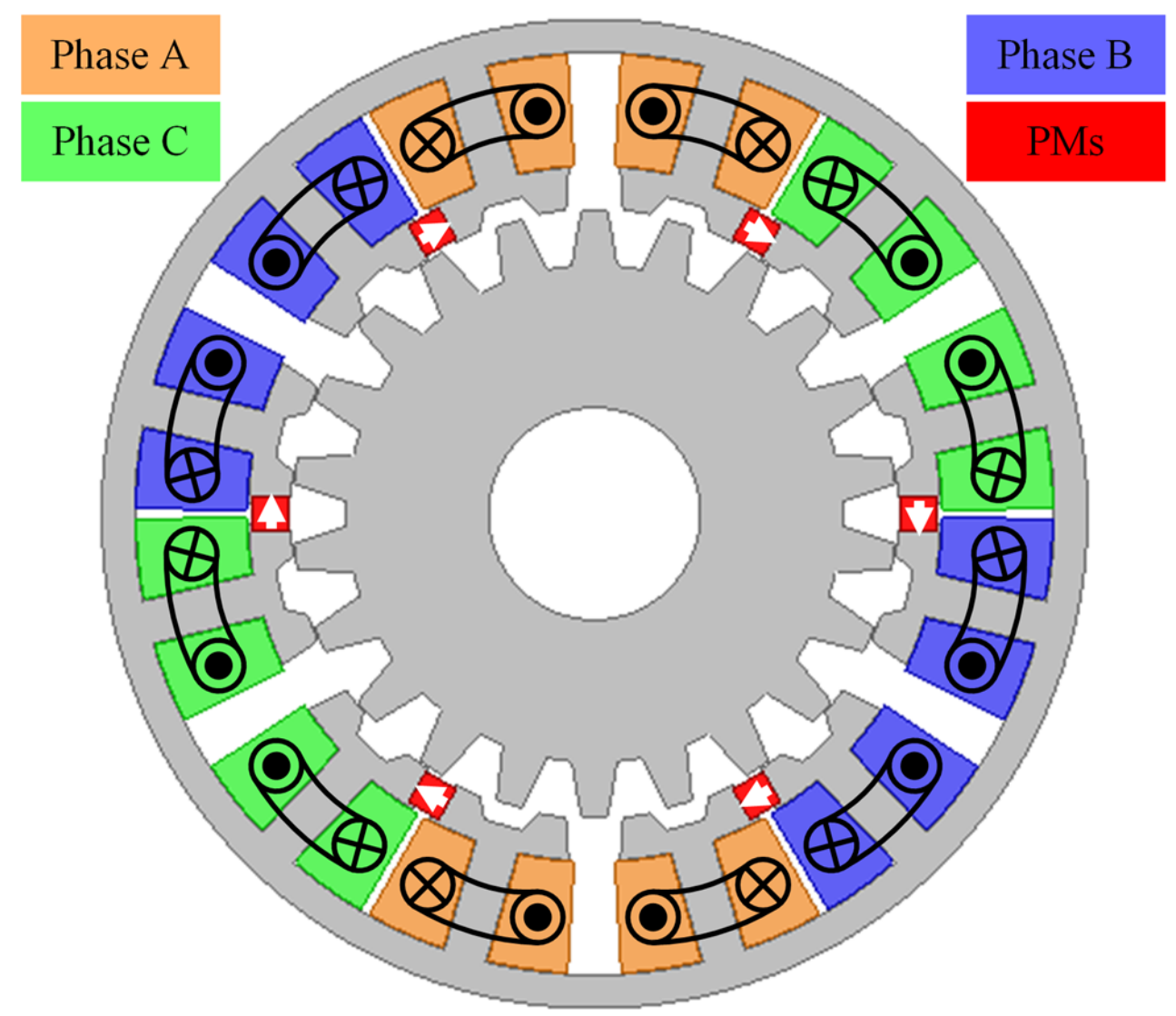

Fig. 1. The topology of the proposed HEMTSRM.

TABLE I. Specifications of Proposed SRM

| Parameter | Value |
|---|---|
| **Stator outer diameter (mm)** | 140 |
| **Stator yoke thickness (mm)** | 4.72 |
| **Stator pole height (mm)** | 16.12 |
| **Air-gap length (mm)** | 0.3 |
| **Rotor pole height (mm)** | 7.64 |
| **Stator tooth arc (deg)** | 4.87 |
| **Rotor pole arc (deg)** | 5.06 |
| **Stack length (mm)** | 20 |
| **PM width (mm)** | 5 |
| **PM length (mm)** | 5 |
| **Number of turns per pole** | 140 |
| **PM grade** | NdFeB-N35 |

## II. Proposed Motor

### *A. Topology*

As the topology and specifications of the proposed HEMTSRM are shown in Fig. 1 and detailed in Table I, each phase comprises two C-cores on opposite sides of the rotor to cancel out radial forces. Increasing the number of teeth per stator pole elevates the developed torque [37]-[38], but there is an optimal configuration of two teeth per stator pole as adopted here. Higher numbers result in a smaller commutation angle, necessitating a higher switching frequency, which increases switching and core losses. Additionally, PMs are embedded between the teeth of the C-core of adjacent phases.

## B. Operating Principles

Fig. 2 illustrates the flux paths due to the PMs and the stator coils. At zero or very low currents, the flux developed by the PMs closes its loop through the stator yoke and does not cross the air gap, so it does not contribute to torque production. At higher stator currents, the C-core approaches saturation levels, causing the PM fluxes to pass through the air gap instead of closing their path through the stator yoke. This contributes to energy conversion, resulting in higher torque. The higher the current, the greater the yoke saturation, leading to an increased contribution of the PMs to energy conversion and higher torque production.

## C. Magnetic Equivalent Circuits and Analytical Design

Fig. 3 demonstrates the use of magnetic equivalent circuits (MEC) to analyze the operating principles of SRMs analytically. For simplicity, this analysis ignores leakage flux, mutual flux, and the effects of saturation [13]-[14]. Assuming phase A is excited, the reluctances of the stator yoke, stator pole, rotor yoke, air-gap, and PMs are represented in Fig. 3 as $R_{sy}$, $R_{sp}$, $R_{ry}$, $R_g$, and $R_{PM}$, respectively. The magnetomotive forces (MMF) of the excited coils and PMs are denoted as $F_e$ and $F_{PM}$. $\varphi_{sy}$, $\varphi_{sp}$, and $\varphi_g$ represent the magnetic flux in the stator yoke, stator pole, and air-gap, respectively. As shown in Fig. 3, since the reluctances of PMs are significantly larger than the reluctances of the iron part of the motors, the following equation is always valid.

$$R_{PM} \gg 2R_{sp} \Rightarrow R_{PM} \gg R_{sp} \tag{1}$$

Since two points, A and A', are short circuits. Hence, a simpler equivalent circuit of the investigated designs can be derived, as depicted in Fig. 4. According to Fig. 4, the fundamental equation (2) can be derived using Mesh-Kirchhoff's law. Furthermore, based on Equation (2) and the fact that the reluctances of the permanent magnets (PMs) are always greater than those of the iron parts of the motors, the following equations are consistently valid.

$$\begin{cases} R_{PM} \gg 2R_{sp} + R_{sy} \Rightarrow R_{PM} \gg R_{sy} \\ 3R_{PM} \gg 2R_{sy} + 4R_{sp} + 2R_g + R_{ry} \end{cases} \tag{3}$$

The magnetic flux equations for Mesh-Kirchhoff's law for the proposed Motor have been derived by solving equation (2).

$$\varphi_1 = -\frac{2}{R_1} \times F_e \tag{4}$$

$$\varphi_2 = +\frac{2R_{sy}}{R_2} \times F_e - \frac{R_3}{R_2 R_{PM}} \times F_{PM} \tag{5}$$

$$\varphi_3 = +\frac{2R_{sy}}{R_2} \times F_e - \frac{R_3}{R_2 R_{PM}} \times F_{PM} \tag{6}$$

$$\varphi_4 = +\frac{2R_{sy}}{R_2} \times F_e + \frac{R_4}{R_2 R_{PM}} \times F_{PM} \tag{7}$$

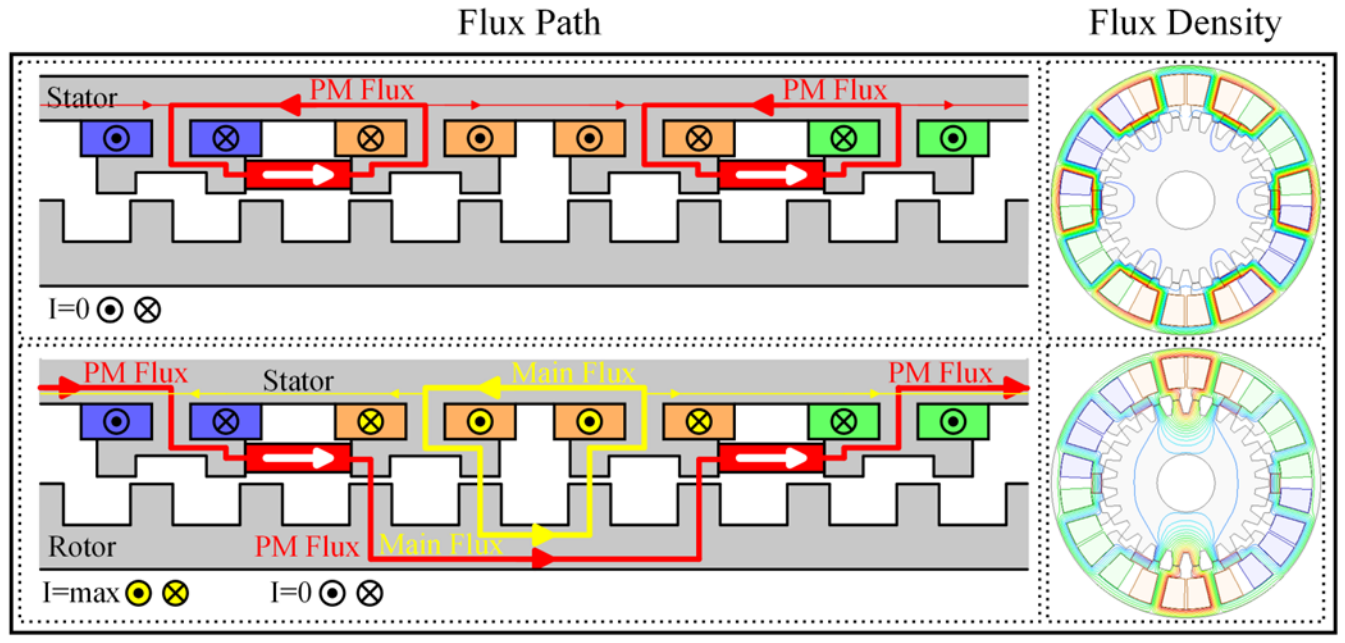

Fig. 2. Flux paths due to PMs and stator coils at zero and maximum currents.

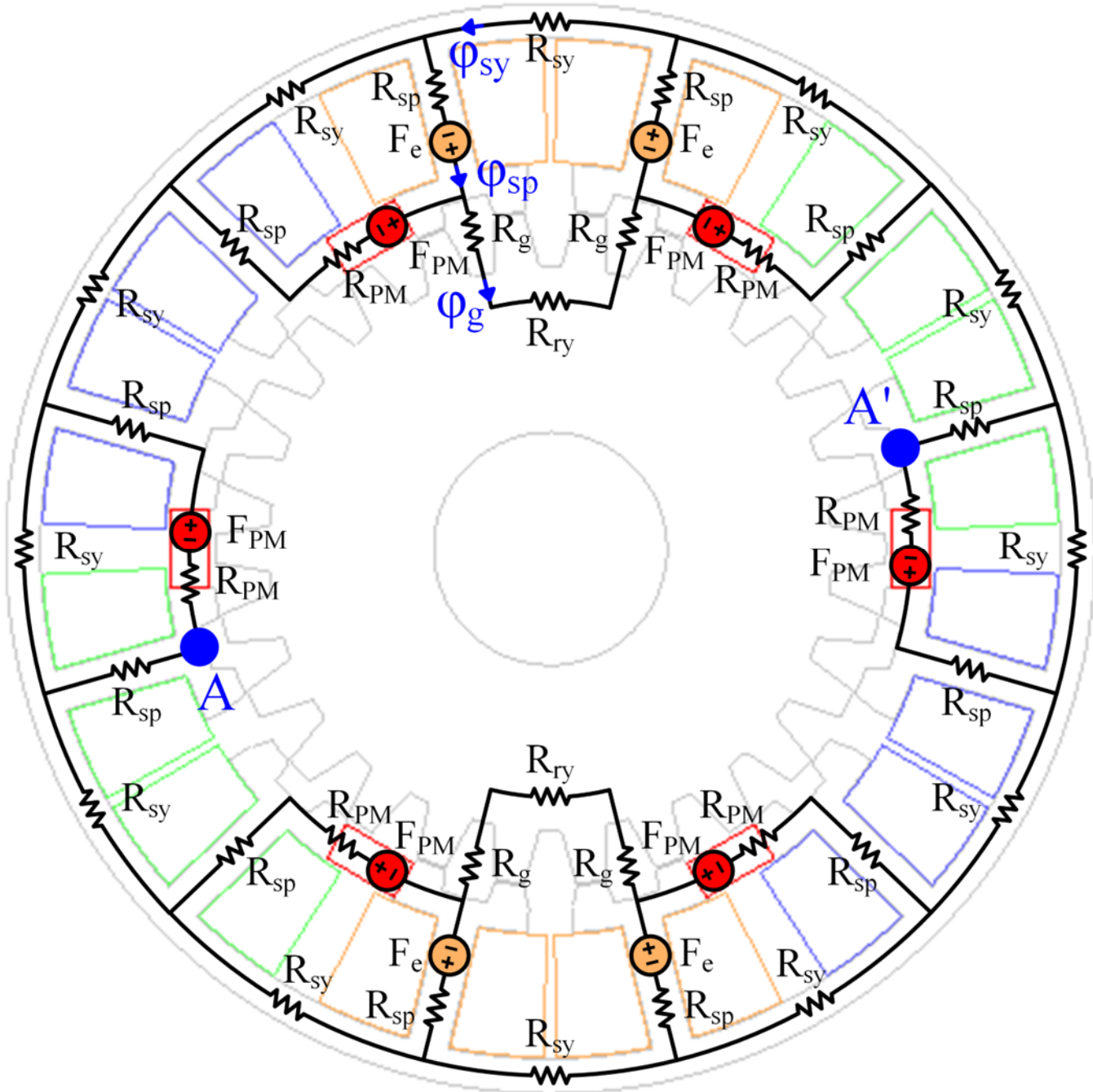

Fig. 3. Magnetic equivalent circuit.

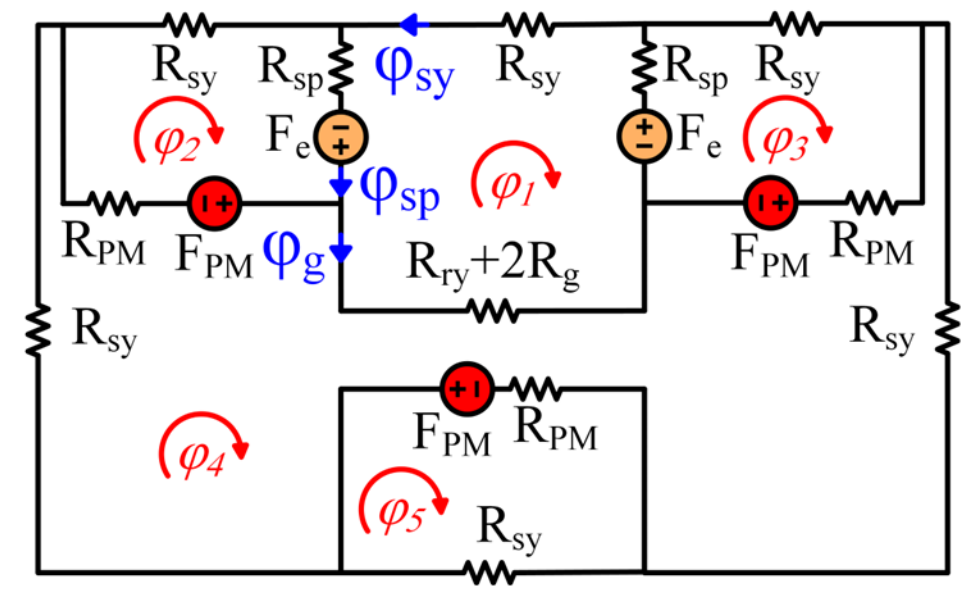

Fig. 4. Simplified magnetic equivalent circuit.

$$\varphi_5 = +\frac{2R_{sy}}{R_2} \times F_e + \frac{R_3}{R_2 R_{PM}} \times F_{PM} \tag{8}$$

in these equations, the following equations are considered:

$$R_1 = \left(2R_g + R_{ry} + 2R_{sp}\right) \tag{9}$$

$$R_2 = \left(4R_g^2 + 4R_g R_{ry} + 8R_g R_{sp} + R_{ry}^2 + 4R_{ry} R_{sp} + 4R_{sp}^2\right) \tag{10}$$

$$R_3 = \left(4R_g^2 + 4R_g R_{ry} + 4R_g R_{sp} + R_{ry}^2 + 2R_{ry} R_{sp}\right) \tag{11}$$

$$\begin{bmatrix} R_{sy}+2R_{sp}+2R_g+R_{ry} & -R_{sp} & -R_{sp} & -2R_g-R_{ry} & 0 \\ -R_{sp} & R_{sy}+R_{sp}+R_{PM} & 0 & -R_{PM} & 0 \\ -R_{sp} & 0 & R_{sy}+R_{sp}+R_{PM} & -R_{PM} & 0 \\ -2R_g-R_{ry} & -R_{PM} & -R_{PM} & 2R_{sy}+3R_{PM}+2R_g+R_{ry} & -R_{PM} \\ 0 & 0 & 0 & -R_{PM} & R_{PM}+R_{sy} \end{bmatrix} \begin{bmatrix} \varphi_1 \\ \varphi_2 \\ \varphi_3 \\ \varphi_4 \\ \varphi_5 \end{bmatrix} = \begin{bmatrix} -2F_e \\ F_e-F_{PM} \\ F_e-F_{PM} \\ 3F_{PM} \\ -F_{PM} \end{bmatrix} \quad (2)$$

$$R_4 = \left(4R_g R_{sp} + 2R_{ry} R_{sp} + 4R_{sp}^2\right) \quad (12)$$

the following equations have been used to derive the magnetic flux equations for the proposed motor's air-gap, stator yoke, and stator pole.

$$\varphi_{sy} = -\varphi_1 ;\ \varphi_{sp} = \varphi_2 - \varphi_1 ;\ \varphi_g = \varphi_4 - \varphi_1 \quad (13)$$

Finally, the equations for the magnetic flux in the air-gap, stator yoke, and stator pole are derived for the proposed motor.

$$\varphi_g = \varphi_4 - \varphi_1 = + \overbrace{\frac{2\left(R_2 + R_1 R_{sy}\right)}{R_1 R_2} \times F_e}^{\varphi'_g} + \overbrace{\frac{R_4}{R_2 R_{PM}} \times F_{PM}}^{\varphi''_g} \quad (14)$$

$$\varphi_{sy} = -\varphi_1 = + \overbrace{\frac{2}{R_1}}^{\varphi'_{sy}} \times F_e \quad (15)$$

$$\varphi_{sp} = \varphi_2 - \varphi_1 = + \overbrace{\frac{2\left(R_2 + R_1 R_{sy}\right)}{R_1 R_2} \times F_e}^{\varphi'_{sp}} - \overbrace{\frac{R_3}{R_2 R_{PM}} \times F_{PM}}^{\varphi''_{sp}} \quad (16)$$

at this motor, phase windings and PMs produce magnetic flux, resulting in $\varphi'_g$, $\varphi'_{sy}$, $\varphi'_{sp}$, $\varphi''_g$, and $\varphi''_{sp}$. As can be seen in equation (14), the magnetic flux within the air-gap experiences an increase as a result of the additional magnetic fluxes $\varphi''_g$. The magnetic flux of PMs contributes to the overall magnetic flux and then permeates into the air-gap. Also, based on equation (15), the magnetic flux of the stator yoke is constant and caused by the magnetic flux produced by phase windings ($\varphi'_{sy}$). According to equation (16), the magnetic flux of the stator pole decreases due to the magnetic flux $\varphi''_{sp}$.

## III. Magnetic Fields, Torque, and Comparisons

Fig. 5 illustrates the distribution of magnetic flux density and flux lines within the optimized HEMTSRM at aligned and unaligned positions under an excitation current of 8 A. The torque-angle characteristics of the proposed SRM for different excitation currents are shown in Fig. 6. In Fig. 7, the components of the total torque, i.e., due to stator current and the PMs, are presented. As expected, at higher currents, the contribution of the PMs to torque production is larger. Mean and peak torque values at different excitation currents are provided in Table II, confirming the aforementioned discussion. Table III compares the proposed HEMTSRM with several existing SRMs and a flux switching permanent magnet (FSPM) motor, highlighting the superiority of the proposed topology in terms of mean torque, torque density (per unit volume), and torque per Ampere. To

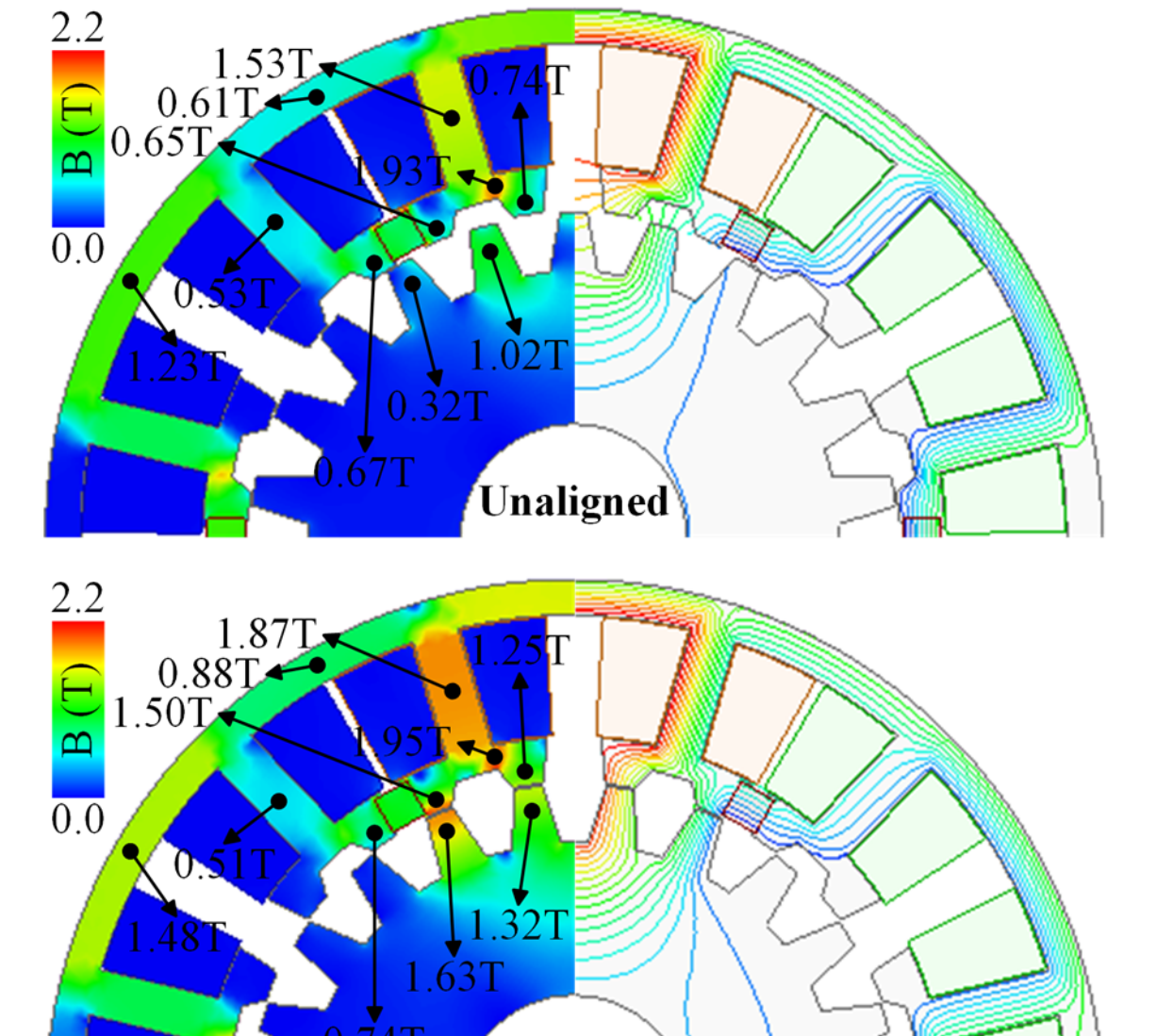

Fig. 5. Magnetic flux density distributions and flux lines.

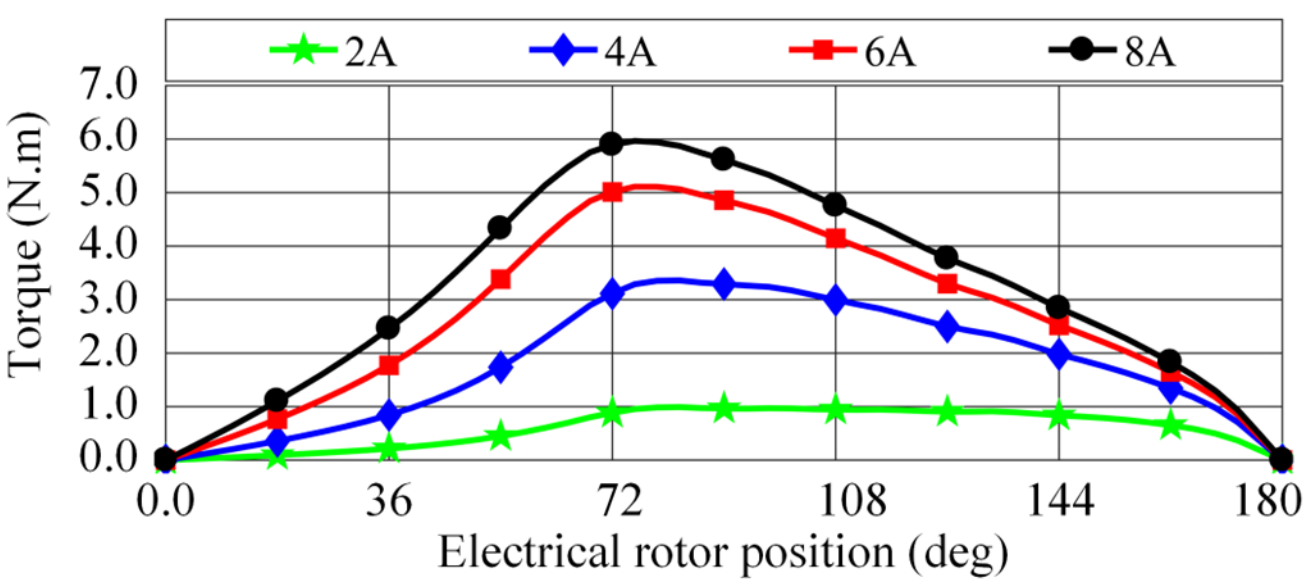

Fig. 6. Torque-angle characteristics of the proposed HEMTSRM.

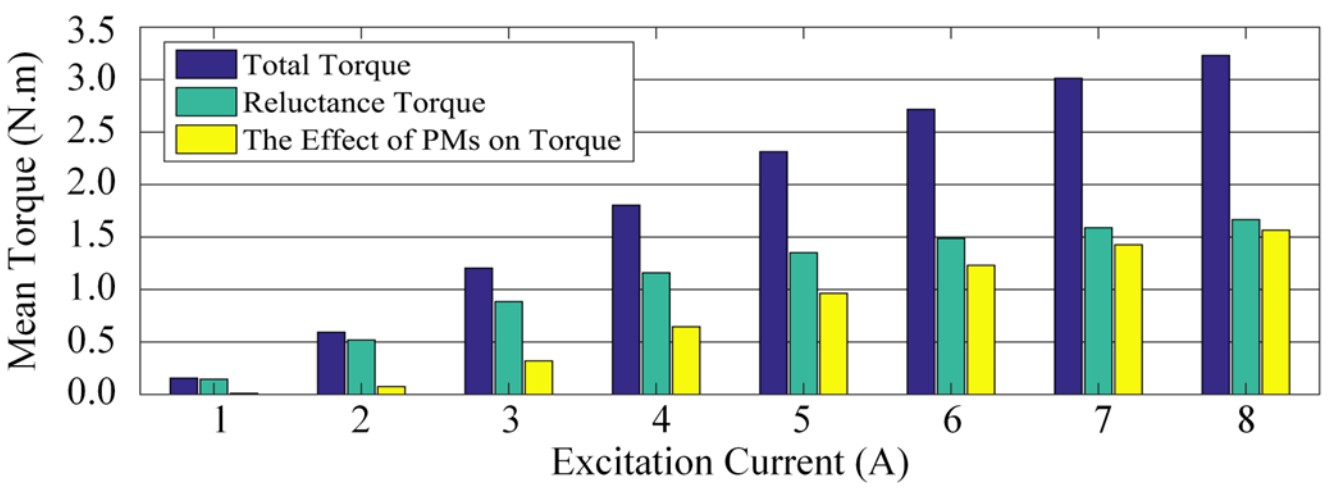

Fig. 7. Mean torque components at different stator currents of the proposed HEMTSRM.

TABLE II. Mean and Peak Torques

| Phase current (A) | Mean torque (N.m) | Peak torque (N.m) |
|---|---|---|
| **1** | 0.156 | 0.258 |
| **2** | 0.594 | 0.990 |
| **3** | 1.205 | 2.137 |
| **4** | 1.805 | 3.359 |
| **5** | 2.315 | 4.397 |
| **6** | 2.718 | 5.110 |
| **7** | 3.015 | 5.607 |
| **8** | 3.232 | 5.967 |

TABLE III. COMPARISON OF THE STATIC RESULTS OF THE PROPOSED 16/18 MHESRM WITH OTHER MOTORS

| Parameter | HEMTSRM | 8/12 SRM | 8/4 SRM | HRM [22] | HRM [31] | MT-HESRM [32] | PM-SRM [32] | HRM [36] | FSPM [39] |
|---|---|---|---|---|---|---|---|---|---|
| **Motor volume with frame (mL)** | 307 | 138 | 138 | 138 | 138 | 138 | 138 | 138 | 138 |
| **PMs volume (mL)** | 3 | - | - | 5.67 | 4.8 | 2.16 | - | 6.21 | 9.45 |
| **Current (A)** | 8 | 6 | 6 | 6 | 6 | 6 | 6 | 6 | 6 |
| **Mean torque (N.m)** | 3.232 | 0.952 | 0.65 | 0.75 | 0.71 | 0.86 | 0.54 | 0.52 | 0.90 |
| **Torque density (N.m/L)** | 10.527 | 6.898 | 4.710 | 5.434 | 5.145 | 6.231 | 3.913 | 3.768 | 6.521 |
| **Torque per ampere (N.m/A)** | 0.404 | 0.158 | 0.108 | 0.125 | 0.118 | 0.143 | 0.090 | 0.086 | 0.150 |
| **Torque density per ampere (N.m/L/A)** | 1.315 | 1.149 | 0.785 | 0.905 | 0.857 | 1.038 | 0.652 | 0.628 | 1.086 |

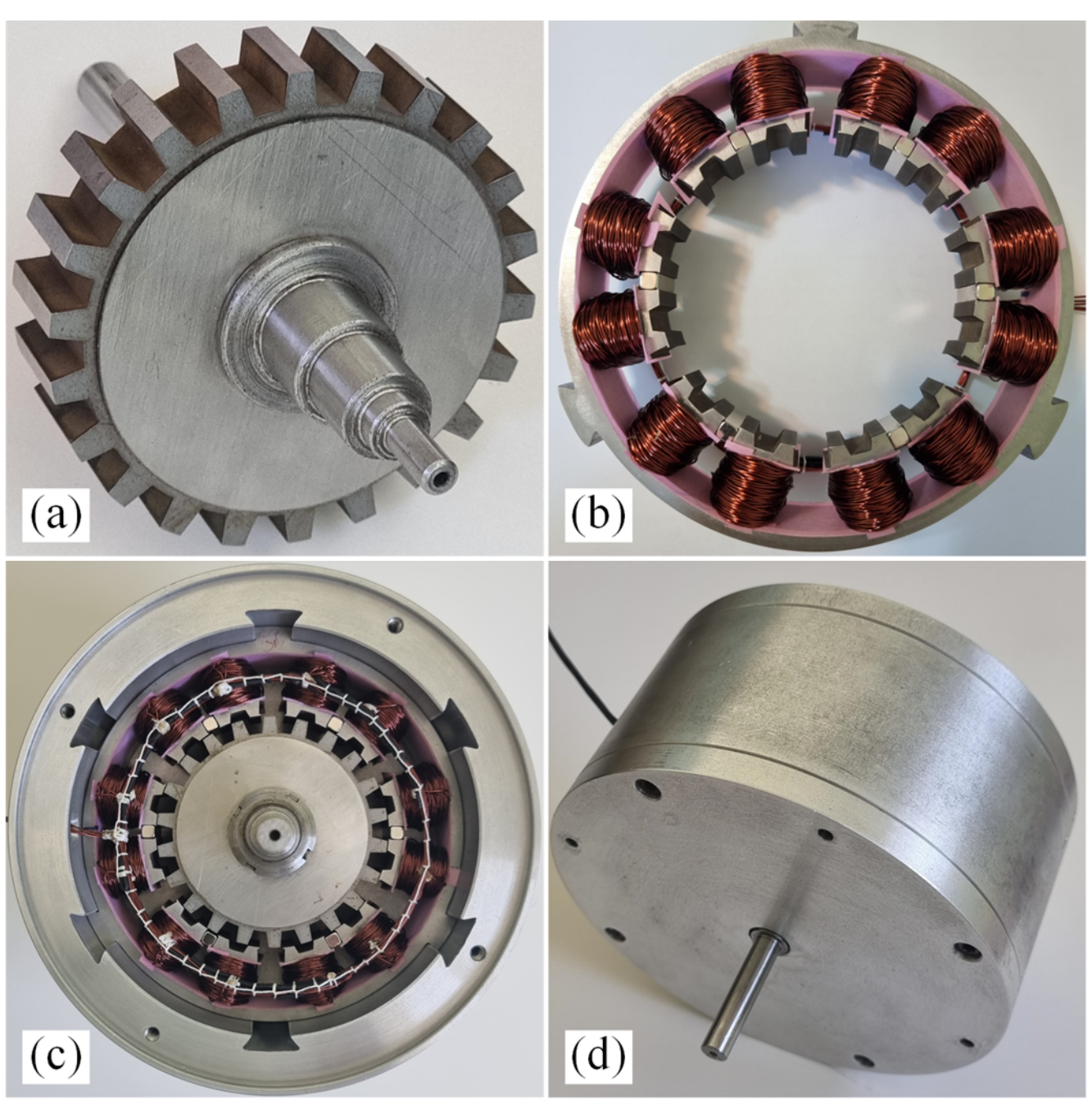

Fig. 8. Prototype of the proposed HEMTSRM: (a) rotor, (b) stator with winding and PMs, (c) assembled motor, and (d) final Motor.

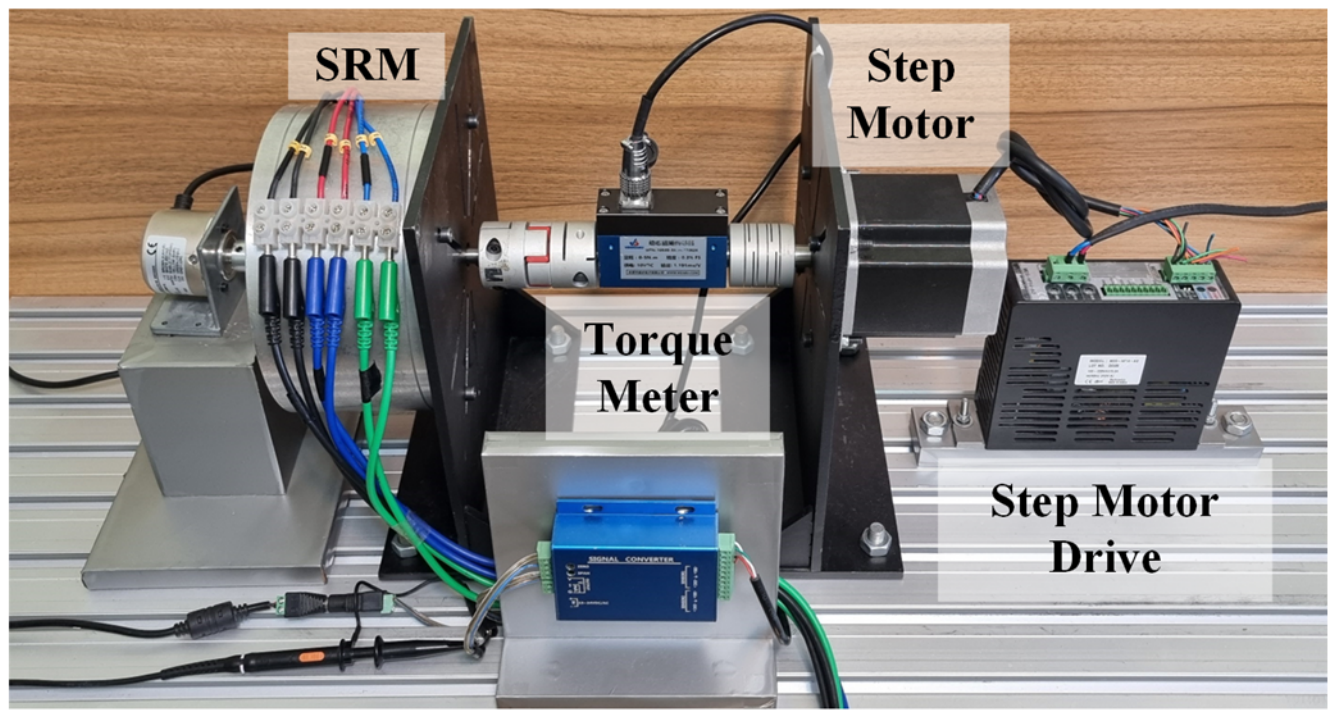

Fig. 9. The test setup.

ensure a fair comparison, the motors are evaluated based on torque density per Ampere, given that their volumes and nominal currents differ, demonstrating the superiority of the proposed HEMTSRM.

## IV. PROTOTYPING AND EXPERIMENTAL RESULTS

The proposed HEMTSRM is prototyped as shown in Fig. 8. As the test setup is shown in Fig. 9, the developed torque at different angles is measured using a stepper motor connected to the SRM through a torque meter. The torque-angle characteristic is given in Fig. 10, representing a great correlation with the results obtained by FEM.

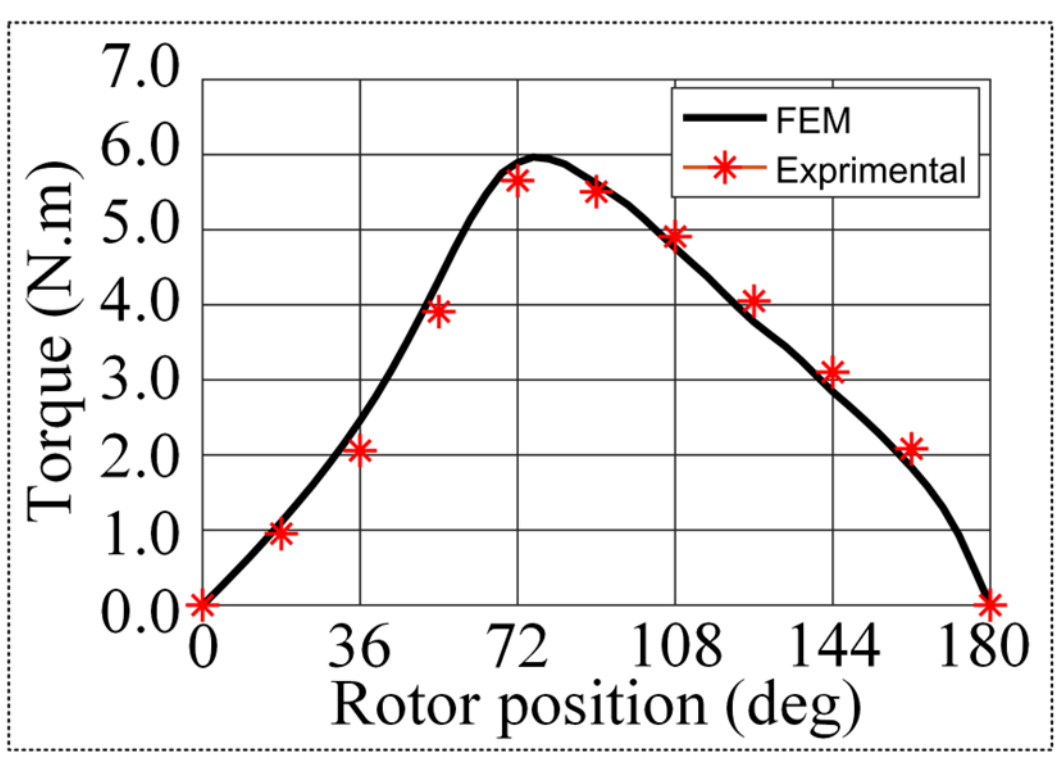

Fig. 10. Measured torque-angle characteristics at 8 A.

## V. CONCLUSION

A three-phase hybrid-excited multi-tooth SRM is proposed, incorporating embedded PMs to improve torque density. This enhancement, combined with the inherent characteristics of SRMs such as robustness, cost-effectiveness, simple manufacturing, and low maintenance, makes it highly suitable for transportation electrification applications. At higher stator currents, where the stator yokes reach saturation, the PMs contribute to energy conversion and torque production. The higher the current, the larger the contribution of PMs to torque production. The operating principles of the proposed SRM are discussed, and the flux density distribution and flux lines within the motor are obtained using FEM. A comparison with a number of conventional switched reluctance motors and a flux switching permanent magnet (FSPM) motor [39] demonstrates the superiority of the proposed SRM in terms of torque density per Ampere. Finally, the proposed motor is prototyped and experimentally tested, confirming the design and the accuracy of FEM results.